\documentstyle[12pt]{article}
\newcommand{\rf}[1]{(\ref{#1})}
\newcommand{\beq}{\begin{equation}}
\newcommand{\eeq}{\end{equation}}
\newcommand{\bea}{\begin{eqnarray}}
\newcommand{\eea}{\end{eqnarray}}

\renewcommand{\l}{\lambda}
\renewcommand{\b}{\beta}

\newcommand{\n}{\nu}
\newcommand{\m}{\mu}

\newcommand{\ep}{\varepsilon}
\newcommand{\vp}{\varphi}

\newcommand{\sg}{\sigma}

\newcommand{\oh}{\frac{1}{2}}
\newcommand{\oq}{\frac{1}{4}}

\newcommand{\dg}{\dagger}
\newcommand{\cO}{{\cal O}}
\newcommand{\Tr}{{\rm Tr}\,}

\begin{document}
\topmargin 0pt
\oddsidemargin 5mm
\headheight 0pt
\headsep 0pt
\topskip 9mm

\addtolength{\baselineskip}{0.20\baselineskip}
\begin{flushright}
NBI-HE-92-20\\
CERN-TH.6508/92
\end{flushright}

\begin{center}

\vspace{36pt}
{\large \bf Parity breaking at high temperature and density}

\end{center}

\vspace{36pt}

\begin{center}
{\sl J. Ambj\o rn } and {\sl K. Farakos}\footnote{Supported by an E.E.C
Fellowship\\
Permanent address:
National Technical University of Athens, Depart. of Physics,GR 157 73
Athens, Greece.}

\vspace{12pt}

 The Niels Bohr Institute\\
Blegdamsvej 17, DK-2100 Copenhagen \O , Denmark\\
and\\
\vspace{24pt}

{\sl M. E. Shaposhnikov}\footnote{On
leave of absence from the  Institute for Nuclear Research of the
Russian  Academy of Sciences,  Moscow 117312, Russia}

\vspace{12pt}

CERN, CH-1211 Geneva 23, Switzerland

\end{center}

\vfill

\begin{center}
{\bf Abstract}
\end{center}

\vspace{12pt}

\noindent
We investigate the question of parity breaking in
 three-dimensional Euclidean $SU(2)$ gauge-Higgs theory by
 Monte Carlo simulations.
 We observe no sign of spontaneous parity breaking
in the behaviour of both local and non-local gauge invariant operators.
However, the presence of parity odd terms in the action can
induce a phase transition to a parity odd ground state which
is characterized by a Chern-Simons like condensate.
The implications for various proposed scenarios of fermion number
non-conservation is discussed.

\vfill

\addtocounter{page}{-1}
\thispagestyle{empty}

\newpage

\section{Introduction}

The non-conservation of the fermion
number in the electroweak theory \cite{hooft}
due to the anomaly in the fermionic current has, in
recent years, attracted a lot of attention.
Under ordinary conditions  the processes
associated with baryon number non-conservation
are exponentially suppressed, since they correspond
to tunnelling between different classical vacua connected by
topological non-trivial gauge transformations.
 However, it has recently  been realized that there can be a great
amplification of anomalous fermion-number non-conservation.
Generally, this might occur when the energy stored in the system
is big enough.  In principle, the energy can be of different forms.
The simplest example is provided by a system at high
temperatures \cite{krs} and/or large fermionic  densities \cite{rub}.
Otherwise, we can consider decays of superheavy fermions \cite{ferm} or
 collisions of particles at high energies \cite{rin,esp}.

It has been  suggested recently to combine non-conservation of the
baryon number with the possibility
of spontaneous parity  breaking \cite{abkw,an,sw,Khl} in
the high temperature limit \cite{ks}
of the electroweak theory and in this way
explain the observed baryon asymmetry of the universe entirely within
the context of the electroweak theory \cite{Shaposhnikov}.
The scenario is as follows.
At temperatures higher than $T_c$, corresponding
to the electroweak transition temperature
$T_c$ ($\sim O(100)GeV$) the $SU(2)$
symmetry is restored, but (by assumption) parity is spontaneously
broken. Since parity is a discrete symmetry
this leads to the creation of domains
with different signs of parity breaking inside
them. However, due to a small but explicit breaking of
$CP$ coming from the KM-matrix,
one type of domain will be energetically favourable, and will eat
the domains with opposite parity.
When the universe has cooled to $T\simeq T_c$
it will be in a state
of maximal parity asymmetry, and
the expectation values of parity odd operators,
which are naturally present in
the high temperature phase, will be different
from zero. Of particular interest  in this connection is the Chern-Simons
condensate, which appears as a term in the effective three-dimensional
high temperature action.  After the electroweak phase transition we
have a situation where the gauge symmetry is broken, but the parity
invariance is restored. The Chern-Simons condensate, characterized by
a non-zero expectation value of the Chern-Simons density $n_{cs}$,
will disappear when parity invariance is restored.
Due to the anomaly of the fermion number current, the  expectation value
$<n_{cs}>$ in the parity broken phase may  be converted to fermions,
thereby explaining the baryon asymmetry observed in the universe.
One additional assumption necessary for the above scenario to be
viable is that the electroweak transition at $T_c$ is first order
\cite{krs,Shaposhnikov}. If it is
 second order, the generated baryon asymmetry
 will be washed away {\it after} the transition, by the very
same baryon number violating processes. The reason is that the effective,
temperature-dependent masses changes smoothly from zero after a
second-order transition.
Consequently, the barrier separation of the different
classical vacua will grow slowly and there will be sufficient time
to create a new thermal equilibrium between baryons and anti-baryons.

\vspace{12pt}

Obviously a large number of assumptions go into the above suggested
mechanism for generation of the
observed baryon asymmetry, and it would be
preferable if one could check some of them.
A determination of the order of the electroweak transition seems
difficult, both from an analytic
point of view \cite{ewbau,chl}, or by means of
lattice gauge simulations\cite{bu}.
The assumption that parity is spontaneously broken
at high temperature is more accessible
to numerical investigation, since the effective
high temperature limit of the static magnetic sector of the electroweak
theory is described by
the three-dimensional $SU(2)$ gauge-Higgs
system \cite{linde2,gpy}.  In the broken phase of the
three-dimensional gauge-Higgs theory it seems
impossible to have spontaneously
broken parity \cite{ks}, but in the unbroken phase, which is the one of
interest in the above cosmological context, infra-red singularities
make a perturbative analysis unreliable. It is the purpose of the
present paper to perform a non-perturbative analysis of the problem
of parity breaking using the technique of lattice gauge theories.

Due to the explicite breaking of parity
and CP-invariance,  the effective action for the gauge and
Higgs fields contains
a number of parity odd terms. The Chern-Simons term has already been
mentioned.
It is given by
\beq
n_{cs} = \frac{1}{16\pi^2} \ep_{ijk} \Tr (F_{ij}A_k-
\frac{2}{3} A_i A_j A_k)  \label{1.0}
\eeq
It appears, for example, in the presence of the non-zero fermionic
number density \cite{action}.
The existence of
a Higgs field leads to other parity odd operators when the fermions
are integrated out.
The simplest operator is
\beq
\tilde{{\cal O}}_1^{cont} = i\ep_{ijk}  F_{ij}^a \vp^\dg \tau^a
\stackrel{\leftrightarrow}{D}_k \vp   \label{1.1}
\eeq
The triangle diagram
leading to this operator is shown in fig.1 and involves
the coupling of two Higgs fields to an $SU(2)$ gauge field.
Clearly, there
is an infinite set of such operators, corresponding to various one-loop
diagrams. A simple example is
\beq
\tilde{{\cal O}}_2^{cont} = i\ep_{ijk} \ep^{abc}
(\vp^\dg \tau^a \stackrel{\leftrightarrow}{D}_k \vp)
(\vp^\dg \tau^b \stackrel{\leftrightarrow}{D}_k \vp)
(\vp^\dg \tau^c \stackrel{\leftrightarrow}{D}_k \vp )  \label{1.2}
\eeq
The pentagon diagram which leads to $\tilde{{\cal O}}_2^{cont}$ in
the high temperature limit is also shown in fig 1.

The coefficients in front of the terms \rf{1.0}-\rf{1.2}
will be extremely small.
They have their origin in the CP-breaking part of the electroweak
theory. Although small, they might still be important if parity
is spontaneously broken, as discussed in \cite{ks}. In the first
part of this article we will mainly consider the terms as small
perturbations to the underlying three-dimensional gauge-Higgs
system, since the main purpose is to find traces of spontaneous
parity breaking and one way to do that is by adding parity-odd
operators and measuring the response. From this point of view
the operators $\tilde{{\cal O}}_1^{cont}$
and $ \tilde{{\cal O}}_2^{cont}$
are in many respects  more convenient  than the
Chern-Simons operator \rf{1.0}. They are, contrary to \rf{1.0},
invariant under local gauge transformation. This means that they have a
natural implementation on the lattice which preserves the parity odd
nature of the terms. There exists no simple and natural lattice
implementation of the Chern-Simons density \rf{1.0}.
Nevertheless \rf{1.1}
and \rf{1.2} have the same origin
in the context of an effective high temperature
expansion of the electroweak theory, and they are       related to the
Chern-Simons term
\rf{1.0} since their sum for a constant Higgs field $\vp_0$
is equal to $n_{cs}$:
\beq
\tilde{{\cal O}}_1^{cont}+\tilde{{\cal O}}_2^{cont} =|\vp_0|^2 n_{cs}
\label{1.3}
\eeq
The use of terms like \rf{1.1} and \rf{1.2} might therefore be superior
to earlier attempts \cite{aaps}
to add directly the Chern-Simons term as a small
perturbation to the effective high temperature lattice action and in this
way test the properties of the vacuum of the three-dimensional theory.
The outcome of these earlier attempts wos ambiguous, and the ambiguity
seems to be related to our inability to find a physical acceptable
representation of the Chern-Simons density on the lattice.

Since there is an infinite number of terms like \rf{1.1} and \rf{1.2},
we will in this article confine ourselves to the study of theories where
only the simplest  source term \rf{1.1} is added. We add it in two
different versions, namely as given by \rf{1.1} and in a ``non-local''
version:
\beq
\tilde{{\cal O}}_3^{cont} =\tilde{{\cal O}}_1^{cont}/|\vp^2|. \label{1.4}
\eeq
The reason for considering $\tilde{{\cal O}}_3^{cont}$ can be found
in eq. \rf{1.3}. In principle we are interested
in adding the Chern-Simons
term as a source term, but as mentioned above we are
unable to do so directly,
and  the non-polynomial interaction
\rf{1.4} might be a good approximation. In the phase where the gauge
symmetry is spontaneusly broken and $\vp$ has only small fluctuations
around a vacuum expectation $\langle \vp_0 \rangle \neq 0$
there should not be any difference\footnote{We use here
the standard continuum notion of spontaneous symmetry breaking. In the
lattice simulation the gauge is not fixed and $\langle \vp \rangle =0$.}
between the source terms \rf{1.1} and
\rf{1.4}. There could be significant differences in the unbroken phase
where $\langle \vp_0 \rangle =0$, and it is our hope that the term
\rf{1.4} would capture a possible difference in this phase between
adding a source term like \rf{1.0} and a source term like \rf{1.1}.

Since the lattice Monte Carlo simulations always use a finite volume,
we cannot strictly speaking have a genuine transition to a parity odd
phase. If spontaneous symmetry breaking is a possibility for the system
at infinite volume, one would nevertheless observe a clear signal
by adding the operator $\tilde{\cO}_1^{cont}$
as a source term, since the expectation
value of $\tilde{\cO}_1^{cont}$
should grow with the volume of the system. We have
illustrated the situation of spontaneous symmetry breaking as compared
to no symmetry breaking in fig.2.
In the case of no spontaneous symmetry  one would expect to
observe a linear growth of
$\langle \tilde{\cO}_1^{cont} \rangle$ as a
function of the coupling strength $\m$,
and almost no volume dependence should be present.
In the following we will report on
the results of MC simulations trying to distinguish between the two
situations of fig.~2.

\section{The phase diagram with $\tilde{\cO}_1^{cont}$ in the action}

For the three-dimensional gauge Higgs system we use the standard lattice
action which is given by:
\bea
S&=& \frac{\b_G}{2} \sum_{x;\m\n} \Tr (1-U_{\Box(x);\m\n}) +
\b_R \sum_x(\oh \Tr \Phi^\dg_x \Phi - v^2)^2+\nonumber \\
 & & \frac{\b_H}{2} \sum_{x; \m} \Tr (\Phi_x-U_{x,x+\m} \Phi_{x+\m})^\dg
 (\Phi_x-U_{x,x+\m} \Phi_{x+\m}).  \label{2.0}
\eea
Here $U_{x,x+\m}\in SU(2)$ denotes the gauge variable associated
with the link $(x,x+\m)$, $U_{\Box(x),\m\n}=
U_{x,x+\m}U_{x+\m,x+\m+\n}U_{x+\m+\n,x+\n}U_{x+\n,x}$ the variable
associated with the lattice point $x$ and the plaquette $\Box$ spanned
by the unit lattice vectors $\hat{\m},\hat{\n}$. Finally $\Phi_x$ denotes
the $SU(2)$ Higgs doublet field associated lattice  point $x$,
but written in matrix form:
\beq
\Phi \equiv
\left( \begin{array}{cc} \phi_1 & -\phi_2^\ast \\
                         \phi_2 & \phi_1^\ast
       \end{array}
\right)
= R \cdot V \mbox{ , }\;\;\; R\in {\bf R}_{+}\mbox{ , }
V\in SU(2) \label{2.1}
\eeq
By taking the  vacuum expectation value $v^2$ as the following function
 of the coupling constants:
\beq
v^2 = \frac{2\b_R+3\b_H-1}{2\b_R} \label{2.2}
\eeq
the naive continuum limit of the theory is obtained
by scaling $\b_R \to 0$
as $a$ while $\b_H \to 1/3$ such that $v^2 \to 0$ as $a$, and \rf{2.0}
goes to the following continuum action:
\beq
{\cal L} = {1\over 4g^2} F^2_{\m \n} +
{|D_\m \phi |}^2 + M_c^2 {|\phi |}^2
+\l_c {|\phi |}^4                                   \label{2.3}
\eeq
The tree value
connection between the lattice parameters in (\ref{2.0}) and the
continuum coupling constants in (\ref{2.3}) is as follows
\begin{eqnarray}
M^2_c & = & \frac{2(1-2\b_R -3\b_H)}{\b_H a^2} \nonumber \\
\l_c & = & \frac{4\b_R}{\b_H^2 a}      \label{2.4}    \\
g^2_c & = & \frac{4}{\b_G a}             \nonumber
\end{eqnarray}

\vspace{12pt}

\noindent
We now add the parity odd source term to the action \rf{2.0}.
First we define lattice quantities corresponding to
$F^a_{\m\n}(x)$ and
$\vp^\dg \tau^a \stackrel{\leftrightarrow}{D}_\m \vp$:
\beq
\frac{i}{2}F^a_{\m\n}(x) \to G^a_{\m\n}(x)=\frac{1}{8} \Tr \tau^a
[ U_{\Box(x);\m,\n}+U_{\Box(x);\n,-\m}+U_{\Box(x);-\m,-\n}
+U_{\Box(x);-\n,\m}]  \label{2.5}
\eeq
\beq
\oh \vp^\dg \tau^a \stackrel{\leftrightarrow}{D}_\m \vp \to
H^a_\m (x)  = \oq \Tr \Phi^\dg_x \tau^a
[ U_{x,x+\m} \Phi_{x+\m}-U_{x,x-\m} \Phi_{x-\m} ] \label{2.6}
\eeq
and the lattice action can then be written as
\beq
S_1= 4\m \sum_{x} \ep_{\m\n\l} H^a_\m (x) G^a_{\n\l} (x) \label{2.7}
\eeq
These lattice terms are chosen in such a way that they have
the correct symmetry, even at the discrete level, and such that
the naive continuum limit of the rhs of \rf{2.5} and \rf{2.6}
agree with the lhs.
The naive continuum limit of the new term is obtained by scaling
$\m \to \m_{c} /a$ while keeping the above-mentioned standard scaling
 of the other couplings. The fact that the continuum coupling
constant in front of a term like $\tilde{\cal{ O}}^{cont}_1$ has the
dimension of length reflects that $\tilde{\cal{ O}}^{cont}_1$ is
not a renormalizable term.

\vspace{12pt}

\noindent
Let us first discuss the phase diagram in the case where
the source term $\m S_1$ is zero. According to our knowledge no detailed
investigation has been made for the model given by \rf{2.0} in three
dimensions. A careful
study of a three-dimensional model where the Higgs
field is in the adjoint
representation was performed in \cite{Nadkarni}.
We found qualitative agreement with the results of ref. \cite{Nadkarni}
for the model \rf{2.0} in the rather large
region of coupling constant space we have scanned.
There are no surprises compared to the standard scenario for a
four-dimensional gauge-Higgs system:  for $\b_G=0$
we have for small $\b_R$ a first-order transition near $\b_H=1/3$,
separating
the broken and unbroken phases. As $\b_R$ increases the $\b_H$ separating
the two phases increases slightly, and the transition becomes
progressively
weaker first order,   ending in a second-order transition at some
finite $\b_H,\b_R$. This picture
is preserved when we move away from $\b_G=0$, and the value
of $\b_R$, which
separates the first- and second order-transition
regions, is  lowered with
increasing $\b_G$. At the other boundary
(i.e. $\b_G=\infty$) the transition
separating the broken and unbroken phase is always second order. When we
move away from $\b_G=\infty$ small values of $\b_R$ will result in a
first-order transition for $\b_H$ close to $1/3$, while larger values
of $\b_R$ result in a second-order transition. In this way there exists
a tricritical line $\tilde{\b}_R(\b_G), \tilde{\b}_H (\b_G)$
extending from the Gaussian fixed point $\b_G=\infty ,
\b_R=0, \b_H=1/3$ and separating the first- and second-order transitions.
We have not located precisely where it ends.

We now discuss the influence of the parity breaking term $ S_1$.
Let us define the  lattice observable
related to $\tilde{\cO}_1^{cont}$ as follows:
\beq
\cO_1 \equiv -\frac{<S_1>}{6\m <R^2>}   \label{def}
\eeq

The procedure has been to fix a value of $\m \neq 0$ and scan
the above phase diagram for the gauge-Higgs system and look for volume
dependence of $<\cO_1>$.
The first statement is that nowhere have we observed
any volume dependence
for $<\cO_1>$. The simulations were performed
on $8^3,12^3,16^3$ and $24^3$
lattices. On the largest lattices it was possible to move quite close
to the gaussian fixed point, but still there was no volume dependence.
We conclude that we have seen no sign of spontaneous parity breaking.

\vspace{12pt}

We do observe, however, explicit parity breaking. No matter what value of
$\b_R, \b_H$ and $\b_G \ne 0$, a sufficiently large value of $\m$
results in a phase transition (or for a large value
of $\b_R$ or $\b_H$ at
least a rapid cross-over) to a state with a large expectation value
of $\cO_1$. The corresponding phase diagram is shown in fig. 3 for
$\b_G=6$ and $\b_R=0.001$. We have three phases denoted
$S$, $SB$ and $PB$.
$S$ denotes a symmetric phase where the expectation value of $R^2$ is
small, the expectation value of the correlator
\beq
\Tr V^\dg U V \equiv \Tr V_x^\dg U_{x,x+\m} V_{x+\m} \label{2.4a}
\eeq
between the phase of the Higgs field and the gauge field is small and
finally the expectation value of $\cO_1$ is small. The phase $SB$ denotes
the phase of spontaneously broken symmetry. It is a continuation of the
similar phase for the ordinary theory with $\m=0$. In this phase
$R^2$ is large, $\Tr V^\dg U V$ is close to two, its maximum value, and
$\cO_1$ is small: there is only a small violation of parity, induced
by the explicit presence of the term with $\m$ in the action. The phase
$PB$ is one where parity appears to be broken in the sense that
the expectation value of $\cO_1$ is large. {\it Note that a positive
expectation value of $\cO_1$, due to the sign convention in \rf{def},
corresponds to a  negative value for the action term $S_1$ given by
\rf{2.3}}. In this phase the expectation
value of $R^2$ is large too, and $\cO_1$ is approximately 0.6 of
its maximal value. The expectation value of $\Tr V^\dg U V$ is small,
as is the expectation value of  $\Tr U_{\Box}$. The three phases meet
at the "triple point" $T$.
We have illustrated the behaviour in fig. 4 of the four observables
mentioned. Deep inside the broken phase (large $\b_H$) the transition
is probably only a rapid cross-over and not a real phase transition.
For the given value of $\b_G$ and $\b_R$ the transition between the
symmetric phase $S$ and the parity broken phase $PB$ is a clear
first-order transition, (it is sharp, independent of the volume and
it is possible to produce a pronounced
hysteresis loop ).
Fig. 5 shows the effect of changing $\m$ for a fixed $\b_H$
which is chosen such that we start in the $S$ phase, afterwards
move into the $SB$ phase and finally end in the $PB$ phase.
Figs. 6a and b illustrate the effect of increasing $\b_R$.
The main effect of such an increase is that
we have to go to somewhat larger $\m$ to induce the
transition to the $PB$ phase. The value of $\m$ reported here is
the value of the $\m_T$ for the triple point where the three phases
meet. The $\b_H$ value for the triple point always remains  smaller than
the $\b_H$ for $\m = 0$ and larger than $1/3$, which is the value
for the Gaussian fixed point corresponding to $\m=0$ and $\b_R=0$.
{}From fig. 6b we have
indications that the triple point moves closer to this gaussian critical
point as $\b_R \to 0$.
Finally Fig. 6c illustrates the behaviour of $\m_T$
versus $\b_R$ and $\b_G$.
We observe first that $\m_T$ tends to a finite value as $\b_G$ tends to
zero for constant $\b_R$ (we will discuss the reason for considering
this particular limit below). On the other hand
$\m_T$ increases with $\b_G$,
but if we want to follow the lines of constant physics
as $\b_G$ increases we must, according to the naive scaling relations,
move to smaller $\b_R$
(recall that $\b_R \sim a$ and $\b_G \sim a^{-1}$).
Doing this we observe that $\m_T$ on the lattice is almost constant,
which indicates that  $\m_{1c}$ approaches zero in a tentative
 continuum limit based on the naive scaling relation. But, as already
mentioned, and as will be discussed in more detail below,
the transition at $\m_T$ is always a first-order transition
and it makes no sense to apply the naive scaling relations.

\vspace{12pt}

Let us try to understand the nature of the new parity
odd phase and examine in more detail the possibility of
associating it with a continuum theory.

{}From fig. 4 the parity broken phase  appears to be a lattice
artifact.  In particular, the plaquette variable $U_{\Box ;\m\n}$
must rotate close to the equator of $S^3$ if we map
$SU(2)$ in the standard
way in the three-sphere, and the smallness of the plaquette
action is {\it not} due to rapid fluctuations of the link variables,
as in usual gauge theories, but is due to an unusual kind of
 alignment with the vector $H^a$. This is clear since $\cO_1$ is
 approximately 0.6 of its maximal value. This
alignment automatically prohibits the usual alignment between the phases
of the gauge field and the Higgs field in the broken phase, and explains
the observed small value of $\Tr V^\dg U V$ in spite of the fact that
the expectation $<R^2 >$ of the Higgs field is large.  The smallness
of $\Tr U_{\Box}$ itself, not only its expectation value, moves any
expansion a long way from the continuum.

 The possibility of a phase transition for large values
of $\m$ is not difficult to understand from a simple mean field argument.
Adding an operator like $\m \tilde{\cO}_1^{cont}$, given by \rf{1.1}
(or $S_1$ given by \rf{2.3} on the lattice)
to the theory corresponds to adding,
in a not too precise way, a tachyonic
mass term to the theory, since it is quadratic in the field $\vp$ and
it will dynamically choose to adjust itself to a
tachyonic coupling, rather
than to a real mass term. (Recall that the expectation value of $S_1$ or
$\mu \tilde{\cO}_1^{cont}$ was negative.)
Of course it has to compete with the real
mass terms present in the theory, but whenever $\m$ is sufficiently large
the tachyonic mode  will have the chance to dominate, provided a suitable
alignment of the gauge fields can be found. By just using the mean field
values for the action terms associated with $\b_G,\b_H$ and $\b_R$ it
is possible to predict the value of $\m$ for which the phase transition
to a "broken parity" phase should occur as a
function of $\b_G,\b_H,\b_R$.
It agrees reasonably with the observed value. Under the assumption that
$\Tr V^\dg U V=0$ and $ U_{\Box}=0$, which is approximately satisfied
deep in the $PB$ phase, we can find a constant field
configuration which seems to agree with the observations.
By a gauge fixing to
the unitary gauge we assume $V_x=1$ and $\Tr V^\dg U V =0$ reduces to
$\Tr U_\m=0$ for all links.  This in combination with $\Tr U_{\Box}=0$
tell us that
\beq
U_{\Box}= i \sg^a u_{\Box}^a,~~~~~~~~~U_\m = i
\sg^a u_\m^a.   \label{mf1}
\eeq
Since $U_{\Box}$ is a product of four $U_\m$'s it is simple algebra to
show that  \rf{mf1} completely fixes the  vectors $u_\m$ corresponding to
links in the 1,2 and 3 directions to satisfy
\beq
       u_\m^a u_\n^a = \pm 1/\sqrt{2}, ~~~~\m\ne \n = 1,2,3   \label{mf2}
\eeq
This in turn implies that
\beq
\frac{1}{6} \ep_{\m\n\l} \;H^a_\m G^a_{\n\l}/{R^2} =
\pm \frac{1}{6} \ep_{\m\n\l} \ep^{abc}u^a_\m u^b_\n u^c_\l =
\pm \sqrt{\sqrt{2}-1}=\pm 0.645 \label{mf3}
\eeq
and the appropriate choice of sign will give a negative value of
$S_1$ which is close to the observed value deep in the $PB$ phase.

Although these mean field arguments are valid deep inside the $PB$-phase,
they are likely to fail close to the transition between the symmetry
phase $S$ and the phase $SB$. For $\m=0$ and
suitable values of $\b_G,\b_H$
and $\b_R$ we have a second-order transition and this transition
seems to persist for small values of $\m$. We have therefore observed
carefully whether there is any sign of enhancement of parity breaking
for a fixed small $\m$, when we move from the symmetric to the broken
Higgs phase (we assume the value of $\m$ so small that we do not have
the phenomenon of tachyonic transition described above).
In fig. 7 we have
shown such a curve for $\cO_1$ as a function of $\b_H$, and we
{\it have} observed an enhancement of the parity breaking transition
precisely at the phase transition. Unfortunately there was still
no volume dependence associated with the observed expectation value of
$\cO_1$, and further the response was linear with $\m$. Again we could
not associate it with any sign of spontaneous parity breaking, and one
possible interpretation of the enhancement is simply that it
 represents a cross-over from one kind of field
configuration, which has only a weak response to the explicit
parity breaking $S_1$ present in the action, namely the configurations
in the unbroken phase, to another, completely different,
kind of configuration, in the broken phase,
which has an equally weak dependence. In between these two well-defined
phases we might get interpolating configurations which
by "accident" trigger
a somewhat larger value of $\cO_1$.

One could ask whether it is possible to use the explicit parity
breaking observed to define a continuum limit.
The first step in that direction is  to locate second-order transitions.
Two locations are possible: in the first case, the transition is reached
from the symmetric
phase of  the  gauge-Higgs system by increasing the "chemical" potential
$\m$, and in the second case it is reached from the broken phase
of the gauge-Higgs system by the same procedure.
A priori the first situation seems much more interesting, since the
symmetric phase is non-perturbative and is infra-red singular from a
perturbative point of view. However, whenever we approached the line of
transitions  from the unbroken phase we observed clearly the phenomenon
of hysteresis, indicating a first-order transition.
It seems unlikely that
this part of the critical line  can be used to
define any continuum theory.
Although it was more difficult to decide whether the transition in the
broken phase was a first- or second-order transition, it is not very
appealing to think of this critical line as one where continuum physics
can be defined. The "perturbative vacuum" where $\m=0$ is certainly
well defined and stable in this phase.
Well into the parity broken phase we have a situation
which differs profoundly from any known continuum assignment of gauge
fields to link variables, and it is hard to imagine the borderline
between these two regimes as being interesting.

One point is singled out, namely the triple point
$T$ where the three phases
meet. For $\m=0$  we expect the transition between the symmetric and
broken Higgs phases to be second order if $\b_G$ is sufficiently large
and $\b_R$ not too small. This feature seems to extend to $\m > 0$, and
it might extend all the way to the triple point $T$
of fig.3, in which case
one could imagine using this second-order point to define a
continuum limit. However, for all the coupling constants we have
checked the situation has been as follows. Even if  the transition
between the phases $SB$ and $S$ starts out as a second order transition
for $\m=0$, and continues to be second-order for $\m$ small,
it always changed to a first-order transition before we reached the
triple point.
We illustrate this situation in fig.8. For $\b_G=5.0$ and
$\b_R=0.01$ we plot $1/2\Tr V^\dg U V $ as a function of $\b_H$ for
two values of $\m$: $\m =0.$ and $\m =0.16$. The second-order transition
at $\m=0$ has changed into a clear first-order one for $\m=0.16$.
Finally  we plot $1/2\Tr V^\dg U V$ and $<\cO_1>$ versus $\m$
for a value of $\b_H$ near the triple point. Again we see a first-order
transition.

We conclude  that we have not been able to find
a second-order phase transition point between $S$ and $PB$.

For larger $\b_H$ we will have a transition between $SB$ and $PB$.
When we move away from the triple point and up in $\b_H$ this transition
is first order, but eventually it seems to change into a cross-over.
Also in this region we found no obvious second-order transition which
could serve as a definition for a continuum limit.

\section{The phase diagram with $\tilde{\cO}_3^{cont}$ in the action}

As mentioned in the introduction we expect the
term $\tilde{\cO}_3^{cont}$
to differ significantly from $\tilde{\cO}_1^{cont}$ in the
symmetric phase $S$. We also expect it to be a better approximation  to a
Chern-Simons like term. For this reason we have repeated the analysis
of the phase structure with this term added instead of
$\tilde{\cO}_1^{cont}$. As the lattice version we have taken
\beq
S_3= 4\m \sum_{x} \ep_{\m\n\l} H^a_\m (x)
G^a_{\n\l} (x)/R^2(x) \label{3.1}
\eeq
The continuum coupling constant $\m_{3c}$ which multiplies
$\tilde{\cO}_1^{cont}$ is dimensionless and if we apply naive scaling,
we have $\m \sim \m_{3c}$ without any lattice scale $a$ entering.

In fig.9 we have shown a phase diagram
similar to the one shown in fig. 3 for two values of $\b_R$.
The qualitative features are the same as in fig.3 except for the
expectation value of $\langle R^2 \rangle$, which seems to be continuous
when we cross from the symmetric phase $S$, to the one of spontaneous
parity breaking $PB$. $\langle R^2 \rangle$ stays small everywhere in
$PB$.

In fig.10 we illustrate the behaviour of our observables as a function
of $\m $. We have chosen $\b_G=6.0$,$\b_R=0.001$  and two values of
$\b_H$ corresponding to a location  below and just
above the triple point.

  Again we see no sign of spontaneous parity breaking for small
$\m$, but a transition for large values of $\m$ to a parity broken
phase $PB$. The transition from $S$ to $PB$ is always first order,
even at the triple point $T$, and the mean field arguments still
apply, since they do not use any properties of $\langle R^2 \rangle$.

The nature of the transition from $SB$ to $PB$ depends on the value of
$\b_R$. For $\b_R=0.001$ it is always first-order even for large values
of $\b_H$, but for $\b_R=0.1$ it starts as a  first order transition
for $\b_H$ just above the value corresponding to the triple point.
For  $\b_H=0.50$ the transition is weakly first order, while for
$\b_H > 0.60$ it has changed into a cross-over.

As for the $\tilde{\cO}_1^{cont}$ term we have to conclude that there
seems to be no second-order phase transition
which can be used for defining
a continuum limit of the parity broken phase.

\section{A "Chern-Simons" condensate}

Although the parity broken phase $PB$ is a lattice artifact, it is still
of some interest to check whether it possesses some of the properties
which were conjectured for the hypothetical continuum version of the
parity broken phase. The most important property was the existence of a
Chern-Simons condensate, which during the phase transition from the
parity broken phase at time $t_1$ to the ``present'' phase of
spontaneously broken
gauge symmetry at time $t_2$ would develop a large value of
\beq
N_{cs}(t_2)-N_{cs}(t_1) = \frac{1}{16\pi^2}
\int_{t_1}^{t_2} dt \int d^3x \;
\Tr \tilde{F}_{\m\n} F_{\m\n}   \label{4.1}
\eeq
where the Chern-Simons number $N_{cs}(t)$ is defined by
\beq
N_{cs}(t) =  \int d^3x \; n_{cs}(x,t). \label{4.2}
\eeq
While $N_{cs}(t)$ itself is not  invariant under topological non-trivial
gauge transformation, the difference given in \rf{4.1}
{\it is} invariant.
Consequently, it makes sense to ask for this value in the case of a
given phase transition (first order, second order etc.)
from the $PB$-phase
to the $SB$-phase. As explained in the introduction we will mainly be
interested in the situation where the electroweak transition is of first
order\footnote{In order to avoid any misunderstanding, let us stress
that the question of a first- or second-order electroweak transition
{\it cannot} be addressed in the present setup.
Since we are working in the
approximation where dimensional reduction is assumed, we cannot ask for
any dynamics associated with temperature. The discussion in the former
sections concerning first- and second-order transitions
only refers to the
possibility of defining a three-dimensional continuum theory from the
lattice theory.}. A first-order transition is well approximated
by a simple change
in the coupling constants \cite{linde}. An
approximation to the first-order
electroweak transition can then be implemented in the following way.
Our starting point is the parity broken phase, which we have reached
from the symmetric phase (corresponding to the high temperatures before
the electroweak transition). We then change the coupling constants
such that they correspond to the broken phase, and relax the gauge
field configuration such that it has a possibility to move to a classical
vacuum configuration in the broken phase. During this motion we measure
the lattice version of the rhs of \rf{4.1} (for a detailed discussion
of this program see \cite{als,aaps}).

One has a large choice of possible relaxation equations.
One possible choice
would be the classical equations of motion with a
damping term (to take away
energy from the hot configuration). Another choice, which we have adopted
here, is to use the simplest relaxation equation which will bring us to
a classical vacuum in the broken phase:
\beq
\frac{\partial \vp(x,\tau)}{\partial \tau} =
-\frac{ \delta S}{\delta \vp(x,\tau)}~,~~~
\frac{\partial A(x,\tau)}{\partial \tau} =
-\frac{ \delta S}{\delta A(x,\tau)}.  \label{4.3}
\eeq
where $S$ is the action for the gauge-Higgs system
and $(A,\vp)$ symbolizes
the gauge- and Higgs fields.

The ``time'' $\tau$ entering in \rf{4.3} is a fictitious relaxation time,
but \rf{4.1} is independent of the choice of
time-parameter since it depends
only on the initial and final field configurations.
Of course the final configuration depends on the specific choice of
dynamics dictated by \rf{4.3}, but since the final configuration is a
trivial vacuum configuration, where only the winding number is of
importance, we do not expect a crucial dependence on the choice of
dynamics during the ``rolling down'' of the Higgs field. At least, the
result of relaxation is invariant to small perturbations of the choice
of relaxation equations since the classical vacua are separated points
of attraction of any relaxation equation.

The measurement of the ``Chern-Simons condensate'',
i.e. of \rf{4.1} according
to the procedure described  above  , is shown in fig.11, both for the
condensate generated from a local parity odd term $\tilde{\cO}_1^{cont}$
and the non-local parity odd term $\tilde{\cO}_3^{cont}$.
We see a clear condensate, where \rf{4.1} grows proportional to the
volume (the density goes to a constant value with volume),
although we have to go to rather large volumes to see
this clearly in the case of $\tilde{\cO}_3^{cont}$.

Fig.12 shows the dependence of the ``Chern-Simons condensate''
on $\m$ for the lattice version of local action $\tilde{\cO}_1^{cont}$
and for constant values of $\b_R ,\b_G$ and $\b_H$.
In the same figure we plot for comparison the value of
$<\cO_1>$ and we see
that the condensate is only formed after the transition to the parity
broken phase.

\section{Discussion}

We have seen no sign of spontaneous breaking of parity.
One could try to discard these lattice results by saying
that the lattices are too small, and that
we are  far from continuum physics. However, from the simulations
of pure $SU(2)$ gauge theories in three dimensions it is known\cite{aop}
that the scaling region is reached already at $\b_G= 5$ and that the
correlation length at this value is quite small. For the pure gauge
theory we can therefore approach continuum physics on
quite small lattices.
Of course the gauge-Higgs system is different and the question of
spontaneous parity breaking a different one. The hope of such a
phenomenon being present is nevertheless born from
the infra-red singularities
present in the gauge-Higgs system in the symmetric phase,
and there is no obvious reason why the
scales involved should be vastly different from the ones which are
believed to control the infra-red singularities of the pure gauge system.

The most natural interpretation of the results is as follows. The
infra-red singularities in three-dimensional non-Abelian gauge theories
(and the magnetic sector of high temperature four-dimensional
non-Abelian gauge theories) can be cured by a dynamical generated
mass scale. We believe such a mass scale is generated for the string
tension\cite{aop} and the glueball mass\cite{ip,farakos}
 in the symmetric phase. This mass scale
could then counteract the potential instability coming from
a Chern-Simons
term. We can write down the following toy model
quadratic effective action
in Euclidean space:
\beq
\int d^3p\; \oh A_k(p)[ (p^2\delta_{ik}-p_ip_k)  + m^2\delta_{ik} +
i\m \ep^{ijk}p_j]A_i(p).
\label{5.1}
\eeq
This action is of course not meant to be
taken seriously, since no reference
is given to the non-Abelian nature of the interaction.
It is meant to serve
as an order of magnitude estimate of  the effects to be discussed.
If $m=0$ the modes with momenta $|p| < \m$ are tachyonic and one
would expect a condensation of the modes $A_i(p)$ which are only
stabilized if we include $A^4$ terms in the effective action.
If $m > 0$ it requires a finite value of $\m$ to trigger a condensation
to a parity broken phase. The above reported results are in qualitative
agreement with the existence of such a mass scale which would prevent any
spontaneous parity breaking. In fact we can, somewhat
courageously, use the
toy model \rf{5.1} to estimate the non-perturbative mass $m^2$ which
enters in the action, since we have determined the
critical $\m$-value where
the tachyonic mode begins to dominate. The most
obvious comparison is made
between the  lattice version of $\tilde{\cO}_3^{cont}$ and \rf{5.1},
since the non-local lattice action is closest to the Chern-Simons like
term in \rf{5.1}. After a little algebra we find that the naive relation
between the $\m$ which appears in \rf{5.1} and $\m_3$ multiplying the
lattice action is given by
\beq
\m \, a = \frac{8\m_3}{\b_G}  \label{5.1a}
\eeq
where $a$ denotes the lattice spacing.
We conclude from the observed critical values of $\m_3$ that
\beq
m \equiv \m = \cO(1)/a   \label{5.1b}
\eeq
The inverse mass seems to be of the order of the lattice spacing.
At first one would be tempted to dismiss this result and simply say
that $\m$ is of the order of the typical lattice excitations, which
have nothing to do with continuum physics. This might be true, but the
remarkable fact is that this very large ``non-perturbative'' mass scale
is precisely what we observe in three-dimensional lattice gauge theories,
whether one likes it or not. It is not our task here to discuss
whether the non-perturbative mass-scales observed in \cite{ip,farakos}
are reasonable or not, we can only conclude that our results are fully
consistent with the non-perturbative
results in \cite{aop,ip,farakos}.

The large value of the inverse magnetic
scale $m_{magn} \sim (2\mu_3)g^2 T \sim 3g^2 T$
introduces large uncertainties in
many quantities connected with the electroweak baryogenesis
scenario. First, any analytical scheme
for the calculation of the finite temperature effective
potential fails for $M_W(\phi) < m_{magn}$
due to the infra-red divergences. This means that we do
not know the shape of the effective potential
at $\phi < 6 g T$, so that it is very difficult
to analyse the dynamics of the phase transition,
for which the behaviour of the potential at small
$\phi$ is essential. Moreover, as has been pointed
out in \cite{Shaposhnikov}, the vacuum expectation
value for the Higgs field should be sufficiently
large in order to suppress B-violation right after
the phase transition, namely $\phi > 3 g T$.
Due to the fact that SU(2) gauge coupling is not so
small,  we cannot trust any analytic
treatment of the scalar effective
potential even for so large $\phi$.
Second, the calculation of the rate of
the fermionic number non-conservation at
temperatures below the phase transition temperatures
relies heavily on the smallness of the
non-perturbative magnetic effects. As we see, our
analysis favours a large magnetic screening mass,
so that there are large uncertainties in the
calculation of the rate too. In other words, two basic
ingredients which can be used for imposing
an upper limit on the Higgs mass from cosmology are
not quite established at the moment.Certainly, much more numerical
work should be done in this direction.

We can now ask whether there are  continuum theories which
have any similarities to the explicit parity broken phase we have
seen on the lattice. The theory of cold neutral fermionic matter
in $(V-A)$ gauge theories considered in \cite{rub} has many similarities.
In the presence of a chemical potential $\m$ for fermions the effective
Hamiltonian of the bosonic fields will essentially be given by the
gauge-Higgs system we have considered in this article plus a Chern-Simons
term where the coupling constant is proportional to $\m$. The Lagrangian
is given by
\beq
-{\cal L}= \frac{1}{2g^2} \Tr F_{\m\n}^2+|D_\m \vp|^2 +
\l (|\vp |^2-\vp_0^2)^2+
\sum_{i=1}^{f} \bar{\psi}_L^{(i)} i\sg_\m D_\m \psi_L^{(i)}.
\label{xx.0}
\eeq
The gauge group is $SU(2)$, $\vp$ is a scalar doublet and
$f$ denotes the number of lefthanded fermions. The  authors in \cite{rub}
studied a situation where there was a fermionic density, neutral
with respect to all gauge charges. Integrating out the fermions led to
the already mentioned Chern-Simons term and they
have shown that in the case where
$\l << g^2$ the normal  vacuum
\beq
\vp =\vp_0,~~~~A_i =0  \label{5.2}
\eeq
is replaced by a more complicated state where
\beq
\vp=0  \label{5.3}
\eeq
and where a gauge field condensate with a Chern-Simons
number different from
zero is created. Our lattice system seems to realize this situation
when coupled to the lattice version of $\tilde{O}^{cont}_3$. The
parity broken phase  $PB$ has the same expectation value of $|\vp|^2$ as
one finds in the symmetric phase,
and a Chern-Simons-like condensate is present.
Ref. \cite{rub} only discusses one of an infinite number of
contributions to the effective Hamiltonian which arises when one
integrates out the fermions from the underlying theory.
In the same way we
have here only considered the lattice version of one of these terms
in the effective action and the final result was a condensate which
was a lattice artifact, but which nevertheless has the essential
features of the condensate suggested in \cite{rub}.

Still another model which exhibits explicit parity breaking is the
three-dimensional {\it Abelian} gauge-Higgs system
 with a Chern-Simons  term\cite{dy}:
\beq
S= -\int d^3x \; |D_\m  \phi |^2+V(\phi)  + \oq \ep^{\m\n\l} A_\m
F_{\n,\l}  \label{xx.1}
\eeq
where $D_\m= \partial_\m -ieA_\m$ and $V(\phi)$ is a Higgs potential.
In the absence of the Maxwell term $\oq F_{\m\n}^2$ the gauge field is
non-dynamical and if one expands around the false vacuum $\phi=0$
there are
just two scalar modes. However, if one expands around the true vacuum
$\phi=\phi_0$ the effective vector action is
\beq
S=- \int  \; \frac{m}{2} A_\m^2 + \oq \ep^{\m\n\l} A_\m F_{\m\n}
\label{xx2}
\eeq
where $m=e^2\phi^2_0$. This first-order system has an alternative
formulation as a conventional second-order system
describing a topological
massive spin 1 particle, with the $A_\m$ essentially the dual of the
Maxwell field strength. Thus the absorbed Higgs scalar has been converted
into an odd parity helicity 1 excitation.

Our model is different from \rf{xx.1} in several aspects.
It is non-Abelian,
it has a ``Maxwell'' term and  it is living in Euclidean space-time.
Nevertheless, it might for extreme values of the coupling constants
have a  parity odd phase where the excitations are of a similar nature.
In order to investigate this possibility we have performed lattice
simulations for very small values of $\b_G$ such that the independent
dynamics of the gauge field is decoupled. However, the conclusion
was the same as for the more conventional values of $\b_G$.
Nowhere we could
find a candidate for a second-order transition between the
$SB$-phase and the
$PB$ phase. The non-Abelian nature of the gauge
group presumably makes the
model radically different from \rf{xx.1}.

\vspace{12pt}

\noindent
{\bf Acknowledgement.}
We thank V. Rubakov and S.Khlebnikov for discussions.
Computer time on an Amdahl VP1100 has been financed by the
Danish Research Council.

\newpage

\addtolength{\baselineskip}{-0.2\baselineskip}

\begin{center} {\large \bf Figure caption} \end{center}

\begin{itemize}
\item[Fig.1]
Triangle diagram which results in the effective term
$\tilde{\cO}_1^{cont}$ given
by \rf{1.1} in the high temperature limit

\item[Fig.2]
(a): $<\cO_1>$ as a function of the coupling strength $\m$ in
the case of spontaneous parity breaking. For a fixed $\m$ we expect
$<\cO_1>$ to grow with the volume of the system. (b): The same functions
as in (a), but with no parity breaking.

\item[Fig.3] The phase diagram for $\b_G=6.0$ and $\b_R=0.001$ and the
action corresponding to $\cO_1$. The used
lattice size is $8^3$. $S$ denotes the symmetric phase,
$SB$ the spontaneously
broken phase, while $PB$ means the parity broken phase.

\item[Fig.4] The behaviour of four observables as a function of $\m$ for
$\b_G=6.0$ and $\b_R=0.001$ for three different values of $\b_H$, which
for $\m=0$ corresponds to the symmetric phase ($\b_H=0.35$, shown with
black dots), just inside the spontaneously broken phase ($\b_H=0.36$,
shown with triangles) and deep inside the broken phase ($\b_H =0.42$,
shown with open circles).

\item[Fig.5]  $<\cO_1>$ (black dots) and $\oh< \Tr V^\dg U V>$
(triangles) for $\b_R=0.001$, $\b_G=6.0$ and $\b_H=0.353$. The
value is chosen in such a way that we start in phase $S$ for $\m=0$. With
increasing $\m$ we move into phase $SB$ as seen from the jump in
$\oh< \Tr V^\dg U V>$, and finally we move into phase $PB$, where
$<\cO_1>$ jumps while $\oh< \Tr V^\dg U V>$ decreases.

\item[Fig.6] (a): The phase transition to $PB$
for various values of $\b_R$.
For a given value of $\b_R$ the open circles correspond to a transition
from $SB$ to $PB$ while the black dots correspond to a transition from
$S$ to $PB$.  (b): The location of the "triple point" $T$ of fig.3
in the $\m,~\b_H$ plane as a function of $\b_R$. The four points
correspond to $\b_R=0.001,~0.005,~0.01$ and 0.05, both $\b_H$ and
$\m$ increasing with $\b_R$. (c): The dependence of $\m_T$ on
$\b_R$ and $\b_G$.

\item[Fig.7]  The expectation value $<\cO_1>$ (open
circles) for a fixed value of
$\m$ as a function of $\b_H$. We are close to $\m=0.0$ and the peak
appears at the transition between the phase $S$ and $SB$, as is seen
from the expectation value of $\oh \Tr V^\dg U V$ (triangles ).

\item[Fig.8] The change in the order of the phase transition from (1):
$S$ to $PB$ as a function of $\b_H$ (fig.8a and fig.8b), and from
(2): $SB$ to $PB$ as a function of $\m$ (fig.8c and fig.8d).

\item[Fig.9] (a): the phase diagram for $\b_G=6.0$
and $\b_R=0.001$ and the
action corresponding to $\cO_2$. (b): the phase diagram for $\b_G=6.0$
and $\b_R=0.01$. The used
lattice size is $8^3$. $S$ denotes the symmetric phase,
$SB$ the spontaneously
broken phase, while $PB$ means the parity broken phase.

\item[Fig.10] Fig.10a:The  expectation value of $R$
as a function of $\m$ for
two different values of $\b_H$ (0.1(circles) and 0.4(dots))
and $\b_R=0.001$, $\b_G=6.0$. The $\b_H=0.1$
data show a transition between
$S$ and $PB$, while the $\b_H=0.4$ data show
a transition between $SB$ and $PB$.
Fig.10b and Fig.10c show $< \cO_3 >$ and $\oh< \Tr V^\dg U V>$ for the
same values of coupling constants.

\item[Fig.11] The measurements of Chern-Simons condensates density as
a function of lattice size for the local and
non-local sources (circles and
dots).

\item[Fig.12] The expectation value of the Chern-Simons condensate versus
$\m$ for $\b_R=0.001$, $\b_H=0.4$ and
$\b_G=9.0$ (circles) and (for comparison)
the expectation value of $< \cO_1 >$ (dots).

\item[Fig.13] Measurement of the
time-integral of $\tilde{F} F$ (eq. \rf{4.1})
during cooling from the parity broken phase to the ordinary phase of
spontaneously broken gauge symmetry. (a) For the local action, (
b) For the
non-local action.

\end{itemize}
\end{document}